\documentclass[conference,a4paper]{APSIPA2023}
\usepackage{multirow}
\usepackage{amsmath}
\usepackage[psamsfonts]{amssymb}
\usepackage{amsxtra}
\usepackage{threeparttable}
\usepackage{graphicx}
\usepackage{booktabs}
\usepackage{float}
\usepackage{array}
\usepackage{float}
\floatstyle{plaintop}

\usepackage{geometry}
\usepackage{fancyhdr}

\fancypagestyle{firststyle} {
    \fancyhf{}
    \fancyhead[C]{2024 Asia Pacific Signal and Information Processing Association Annual Summit and Conference (APSIPA ASC)}

}

\begin{document}

\title{Speech Separation using Neural Audio Codecs with Embedding Loss}

\author{
\authorblockN{
Jia Qi Yip\authorrefmark{1}\authorrefmark{2}
Chin Yuen Kwok\authorrefmark{1}
Bin Ma\authorrefmark{2} and 
Eng Siong Chng\authorrefmark{1}
}

\authorblockA{
\authorrefmark{1}
Nanyang Technological University, Singapore\\}

\authorblockA{
\authorrefmark{2}
Alibaba Group \\
E-mail: jiaqi006@e.ntu.edu.sg}
}

\maketitle
\thispagestyle{firststyle}
\pagestyle{fancy}

\begin{abstract}
Neural audio codecs have revolutionized audio processing by enabling speech tasks to be performed on highly compressed representations. Recent work has shown that speech separation can be achieved within these compressed domains, offering faster training and reduced inference costs. However, current approaches still rely on waveform-based loss functions, necessitating unnecessary decoding steps during training. We propose a novel embedding loss for neural audio codec-based speech separation that operates directly on compressed audio representations, eliminating the need for decoding during training. To validate our approach, we conduct comprehensive evaluations using both objective metrics and perceptual assessment techniques, including intrusive and non-intrusive methods. Our results demonstrate that embedding loss can be used to train codec-based speech separation models with a 2x improvement in training speed and computational cost while achieving better DNSMOS and STOI performance on the WSJ0-2mix dataset across 3 different pre-trained codecs.
\end{abstract}

\section{Introduction}
Speech separation, also known as the cocktail party problem~\cite{cherry1953some}, is the task of isolating individual speakers from overlapping audio, has seen remarkable progress in recent years~\cite{sepformer}. While current models achieve impressive signal-to-noise ratios on clean datasets~\cite{SPGM}~\cite{zhao2023mossformer2}, the field is now advancing towards more generalized capabilities in diverse environments. This shift towards broader applicability necessitates even larger training datasets~\cite{pons2024gass}, a requirement that is currently hindered by the substantial computational costs associated with training speech separation models.

The computational intensity of speech separation stems from two primary factors. First, traditional models typically employ minimal time-compression in their encoders, often using simple 1D convolutions that achieve only 8x downsampling in the time dimension~\cite{Subakan2022ExploringSM}. Recent approaches have revisited hybrid frequency and time domain methods~\cite{Wang2022TFGridNetIF}~\cite{chen2023neural}, incorporating greater downsampling to enhance computational efficiency without sacrificing performance.

Second, the loss function and evaluation metrics used in speech separation are inherently expensive. The Permutation Invariant Training (PIT)~\cite{PIT} Loss, essential for handling the speaker order ambiguity, requires computing $O(N!)$ combinations of outputs and ground truth, where N is the number of speakers in the mixture. Furthermore, the commonly used Scale-Invariant Signal to Distortion Ratio (SI-SDR)~\cite{le2019sdr} is a waveform comparison loss function, which is computationally demanding for the long sequences typical in audio waveforms.

Neural Audio Codecs~\cite{zeghidour2021soundstream}~\cite{encodec}~\cite{DACkumar2024high}, self-supervised models trained on vast amounts of general audio data, offer a promising solution to these challenges. These codecs, structured as autoencoders with a quantizer~\cite{Gray1984VectorQ} between the encoder and decoder, can achieve time-domain compression ratios of approximately 100x, depending on the specific codec architecture. Due to the strong representations learned by neural audio codecs, they have been found to have useful applications beyond compression. They have been utilized in various downstream tasks, including automatic speech recognition~\cite{wang2023viola}~\cite{gupta2024exploring}, text-to-speech~\cite{junaturalspeech}~\cite{yang2024simplespeech}, speaker verification~\cite{puvvada2024discrete}, and singing voice synthesis~\cite{chang2024interspeech}. Recently, speech separation has also been added to this list of applications~\cite{yip2024towards}.

\begin{figure}[t!]
  \centering
  \includegraphics[width=\linewidth]{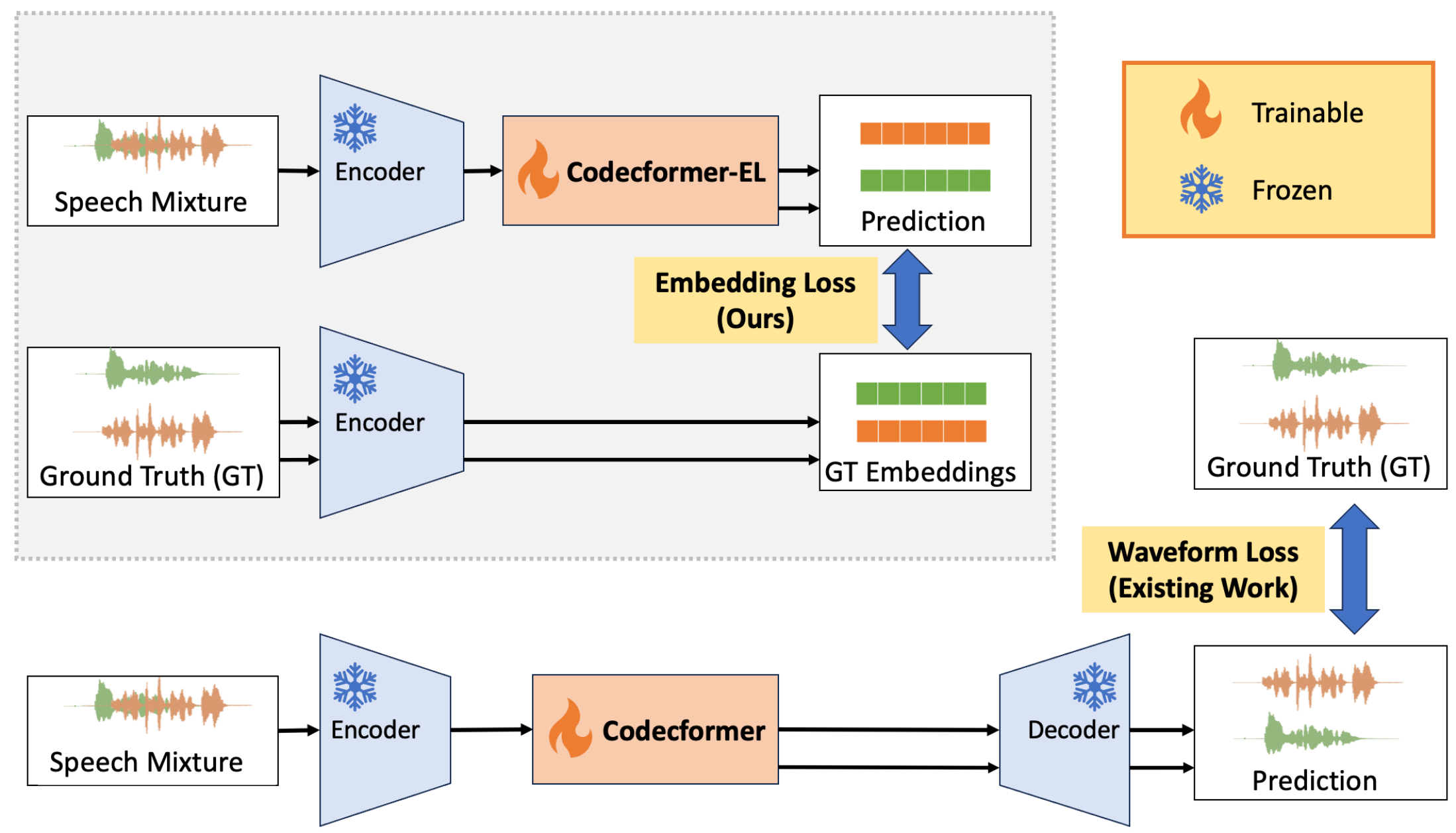}
  \caption{Comparison between our proposed embedding loss (Top Left) with the conventional waveform loss used in a previous codec-based speech separation approach (Bottom). The key advantage of the embedding loss training method is that the decoder of the neural audio codec is not necessary for training, significantly reducing the computational cost during training.}
  \label{fig:overview}
  \vspace{-5pt}
\end{figure}

The recently proposed Codecformer~\cite{yip2024towards}, a neural audio codec-based speech separation model, has demonstrated the potential of integrating neural audio codecs with speech separation models to reduce computational requirements. However, as shown in Figure~\ref{fig:overview}, this approach still relies on waveform-comparison losses, leaving room for further optimization.

In this paper, we propose a novel embedding loss for speech separation using neural audio codecs as shown in Figure~\ref{fig:overview}. Our approach operates directly on the encoder representations of the codec, eliminating the need for expensive waveform comparisons during training. While PIT is still necessary, the shorter sequence length of the embeddings substantially decreases the computational cost of loss calculation.

\noindent Our contributions are threefold:
\begin{itemize}
\item We demonstrate that neural audio codec-based speech separation models can be trained from a variety of codecs using only embedding level loss, resulting in 2x improvement in training speech and computational cost with better performance on perceptual metrics, despite lower objective scores.

\item While Codecformer~\cite{yip2024towards} only supported the Descript Audio Codec (DAC)~\cite{DACkumar2024high} and the PESQ~\cite{rix2001perceptual} perceptual quality metric, we expand support to include EnCodec and SoundStream, as well as include additional perceptual quality metrics, STOI~\cite{taal2011algorithm} and DNSMOS~\cite{reddy2021dnsmos}.

\item We investigate the impact of pre-training data quantity and diversity on separation performance, comparing separation using the DAC model trained on open source datasets, LibriTTS, AMUSE, and the original DAC~\cite{DACkumar2024high} dataset.

\end{itemize}
\vspace{-5pt}
\section{Methodology}

In this work we propose Codecformer-EL, based on the recently proposed Codecformer~\cite{yip2024towards}, which modifies the training method for Codecformer by employing embedding loss as shown in Figure~\ref{fig:overview}. While Codecformer utilizes both the encoder and decoder of the neural audio codec and performs waveform comparison loss, Codecformer-EL only requires the encoder. The loss computed is mean squared error (MSE) loss wrapped within the PIT~\cite{PIT} algorithm.

As shown in Figure~\ref{fig:overview}, we perform speech separation by training a separator model, Codecformer~\cite{yip2024towards}, that makes use of a frozen encoder and decoder from a pre-trained neural audio codec. During inference and evaluation, the decoder is used to generate the necessary waveform for comparison.

\subsection{Embedding Loss Function}
Our proposed Embedding loss function can be written as follows:
\begin{equation}
\begin{aligned}
    e &= \text{Encoder}(s_{gt}) \\
    \text{MSE} &= \frac{1}{n}\sum_{i=1}^{n}(\hat{e}_{i} - e_{i})^{2}
   \end{aligned}
\end{equation}

where $s_{gt}$ is the ground truth speech and $\hat{e}$ is the separated embeddings produced by the separator in Codecformer-EL and $n$ is the number of samples in each batch. This loss function is only used during training. Since the encoder used here to produce the groundtruth embeddings, $e$, is frozen, the embeddings can either be generated during training to save memory, or can be pre-computed before training to save on compute.

The choice of mean squared error loss in this study is inspired by early speech separation work that used this loss over spectrograms. However, because the neural audio codecs are learned representations, the embeddings obtained by the encoder are not spectrograms. Nevertheless, these embeddings are representations of audio and can be decoded into audio. Thus they may be thought of as spectrogram-like, which makes MSE loss more appropriate as opposed to Cosine Similarity loss used for semantic representations. 

\begin{figure}[t!]
  \centering
  \includegraphics[width=\linewidth]{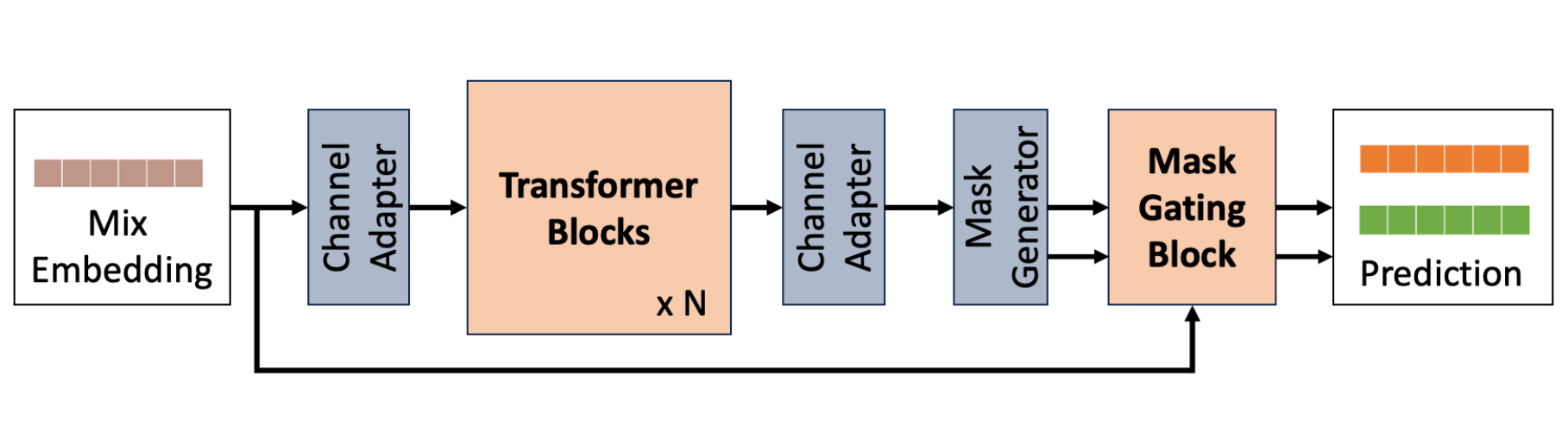}
  \caption{Overview of the Codecformer model based on~\cite{yip2024towards}. The bulk of the model consists of N transformer blocks along with channel adapter layers, a mask generator and the mask gating block.}
  \label{fig:modelArch}
\end{figure}

\subsection{Neural Audio Codecs}
To verify that the embedding loss proposed in this paper is applicable to neural audio codec-based speech separation more generally, we make use of three popular neural audio codecs in this study, SoundStream~\cite{zeghidour2021soundstream}, EnCodec~\cite{encodec} and Descript Audio Codec (DAC)~\cite{DACkumar2024high} as described in Table~\ref{tab:CodecComparisons}. We also compare the differences in performance across different training datasets, LibriTTS~\cite{zen2019libritts}, AMUSE~\cite{ESPnetcodec}, and the original DAC dataset for the DAC model. All the codecs with embedding size of 512 in Table~\ref{tab:CodecComparisons} were pre-trained using ESPnet-Codec~\cite{ESPnetcodec}.

\renewcommand{\arraystretch}{1.2}
\begin{table}[h!]
  \caption{Comparison of the different codecs used in this work}
  \label{tab:CodecComparisons}
  \centering
    \begin{tabular}{ |c|c|c|c|  }
    \hline
    Model & Pre-training Dataset & Embedding Size & Params (M) \\
    \hline
    SoundStream & AMUSE & 512 & 19.7\\
    EnCodec & AMUSE & 512 &  19.7\\
    DAC & LibriTTS, AMUSE & 512 & 19.7\\
    DAC & Original & 1024 &  74.2 \\
    \hline
    \end{tabular}
\end{table}

\noindent \textbf{SoundStream}~\cite{zeghidour2021soundstream} pioneered the use of Residual Vector Quantization (RVQ) in neural audio codec development. Its architecture combined a SEANet-based encoder-decoder with an RVQ quantizer. For discrimination, it employed both waveform and spectral discriminators. The model's loss functions included adversarial and feature-matching losses from these discriminators, along with a multi-resolution mel-spectrogram loss.

\noindent \textbf{EnCodec}~\cite{encodec} built upon SoundStream's design, substituting the STFT discriminator with a multi-scale version. It introduced a probabilistic approach to discriminator updates and implemented a loss balancing mechanism to automatically adjust loss scales during training.

\noindent \textbf{DAC}~\cite{DACkumar2024high} further refined EnCodec by incorporating several enhancements. These included the adoption of the snake activation function, advanced RVQ featuring factorized and L2-normalized codes, increased use of quantizer dropout, and implementation of an MSMPMB discriminator. These improvements aimed to tackle codebook collapse, the phenomenon where a fraction of the codes are unused.

The LibriTTS dataset~\cite{zen2019libritts} is a commonly used for training neural audio codecs. The Audio, Music, and Speech Ensemble (AMUSE) dataset developed for ESPnet-Codec~\cite{ESPnetcodec}, is a fused corpus consisting of multiple open source high quality audio data, which also includes the original DAC dataset. Thus, the models trained on the AMUSE dataset represents the models with the largest-scale pre-training.

\subsection{Codecformer-EL}
The Codecformer with embedding loss (Codecformer-EL) model proposed in this study makes use of the same basic architecture as Codecformer~\cite{yip2024towards} as shown in Figure~\ref{fig:modelArch}. The bulk of the model consists of a stack of N transformer blocks followed by a mask gating block that is responsible for modifying the mix embedding received from the encoder. The Channel Adapters are linear layers that adapt the codec embedding size to the required dimensions of the transformer blocks. The mask generator is a linear layer that increase the number of channels to match the number of speakers in the mixture.

One weakness of the original Codecformer~\cite{yip2024towards} it that the model only supported the DAC encoders and decoders. In this work we adapted Codecformer to support EnCodec and SoundStream as well. This required modifying the activation function in the Mask Gating Block from the Snake activation function used in DAC, to the ELU~\cite{clevert2015fast} activation function used in EnCodec and SoundStream. By matching the activation function in Codecformer with that of the neural audio codecs that they are embedded in, the generated outputs of the separation model are more likely to adopt a distribution more similar to the distribution expected by the decoder, potentially resulting in better generation by the decoder.

\section{Experiments}
\subsection{Dataset}
Our experiments utilize the widely-used WSJ0-2mix benchmark dataset~\cite{wsj0mix}. This dataset is derived from the WSJ0 corpus, which consists of English read speech from the Wall Street Journal. The dataset comprises speech mixtures created from clean audio recordings. All audio, including both the mixtures and their corresponding ground truth, is sampled at 8kHz and recorded in acoustically controlled settings. The synthetic mixtures are generated using the 'min' condition, where the length of each mixture is cropped to match the duration of the shorter of the two input speeches. The WSJ0-2mix dataset is structured as follows: A training set consisting of 20,000 mixtures (30 hours) from the WSJ0 si\_tr\_s set, validation set consisting of 5,000 mixtures (10 hours) from the WSJ0 si\_dt\_05 set and a test set consisting of 3,000 mixtures (5 hours) from the si\_et\_05 set.

\subsection{Codec Implementation}
To ensure a consistent implementation, the model code and pre-trained weights for the codecs used were obtained from the ESPnet-Codec repository~\cite{ESPnetcodec} except for the DAC Original implementation where the code and weights were obtained from the official release of the original authors\footnote{https://github.com/descriptinc/descript-audio-codec}. Meanwhile, the training and evaluation of the models was implemented on the speechbrain toolkit~\cite{speechbrain} with the models from ESPnet handled by a wrapper written for speechbrain. For all neural audio codec models, the 16kHz version of the model is used, which is the lowest sampling rate available across all models. To manage the difference in sampling rate of the dataset and the neural audio codecs, the input audio is resampled using the torchaudio resampling method. Although this means that the codec receives audio that has missing frequencies, it is not expected that this results in significant performance impact.

\subsection{Evaluation Methods}
In addition to the standard objective metrics used for speech separation, such as SI-SDR, SI-SDRi, SDR, SDRi, we make use of perceptual evaluation metrics to measure the performance of the models. The perceptual evaluation metrics are necessary when performing neural audio codec-based speech separation because distortions introduced by the codecs result in distortions that are heavily penalized by objective metrics but are do not affect perceptual quality.

The perceptual quality metrics used in this study are deep noise suppression mean opinion score~(DNSMOS)~\cite{reddy2021dnsmos}, perceptual evaluation of speech quality~(PESQ)~\cite{rix2001perceptual} and short-time objective intelligibility~(STOI)~\cite{taal2011algorithm}. Among these metrics, DNSMOS is a non-intrusive metric that does not rely on reference speech, while PESQ and STOI are intrusive metrics that do rely on some reference speech. Due to the ambiguity in the output order of the model, the ideal permutation for each of these metrics is computed based on maximizing the SI-SDR over all possible permutations. The permutation of outputs against the ground-truth that produces the highest average SI-SDR is used to compute all of the perceptual quality metrics. This also ensures that a consistent permutation is chosen across all objective and perceptual evaluation metrics.

\renewcommand{\arraystretch}{1.2}
\begin{table*}[t!]
  \caption{Comparison of Separation performance on different Neural Audio Codec Models trained on embedding and waveform losses. All codec models were pre-trained on the AMUSE dataset}
  \label{tab:modelComparisons}
  \centering
    \begin{tabular}{ |p{3cm}|c||c|c|c|c||c|c|c|c|c|c|  }
     \hline
    \multirow{2}{*}{\textbf{Model}} & \multirow{2}{*}{\textbf{Loss Type}} & 
    \multicolumn{4}{c||}{\textbf{Objective Metrics}} &
    \multicolumn{4}{c|}{\textbf{DNSMOS}} &
    \multicolumn{1}{c|}{\multirow{2}{*}{\textbf{PESQ}}} & 
    \multicolumn{1}{c|}{\multirow{2}{*}{\textbf{STOI}}}  \\ \cline{3-10} 

     \multicolumn{1}{|c|}{} & \multicolumn{1}{c||}{} & \multicolumn{1}{c|}{\textbf{SI-SDR}} & \multicolumn{1}{c|}{\textbf{SI-SDRi}} & \multicolumn{1}{c|}{\textbf{SDR}} & \multicolumn{1}{c||}{\textbf{SDRi}} & \multicolumn{1}{c|}{\textbf{OVRL}} & \multicolumn{1}{c|}{\textbf{SIG}} & \multicolumn{1}{c|}{\textbf{BAK}} & \multicolumn{1}{c|}{\textbf{p808}} & \multicolumn{1}{c|}{} & \multicolumn{1}{c|}{} \\ \hline

    DAC & Embedding & -1.2 & -1.3 & 1.1 & 0.9 & 1.80 & \textbf{2.12} & 2.99 & \textbf{2.58} & 1.80 & \textbf{0.81} \\
    DAC & Waveform & 2.8 & 2.8 & 4.9 & 4.7 & \textbf{1.81} & 2.04 & \textbf{3.38}& 2.55 & \textbf{2.06} & 0.79 \\
    \hline
    EnCodec & Embedding & -29.1 & -29.1 & -8.5 & -8.6 & \textbf{1.89} & \textbf{2.20} & \textbf{3.18} & \textbf{2.65} & \textbf{1.92} & \textbf{0.80} \\
    EnCodec & Waveform & -9.0 & -9.0 & -2.6 & -2.8 & 1.24 & 1.26 & 2.75 & 2.29 & 1.28 & 0.51\\
    \hline
    SoundStream & Embedding & -14.2 & -14.2 & -4.9 & -5.00 & \textbf{1.80} & \textbf{2.07} & 2.98 & \textbf{2.63} & \textbf{1.94} & \textbf{0.83}\\ 
    SoundStream & Waveform & -2.5 & -2.4 & 1.2 & 1.0 & 1.73 & 1.95 & \textbf{3.35} & 2.45 & 1.77 & 0.74 \\
    
    \hline
    \end{tabular}
    
\end{table*}
\renewcommand{\arraystretch}{1.2}
\begin{table*}[t!]
    \vspace{10pt}
  \caption{Comparison of codecformer separation performance on the DAC model with different pre-training datasets}
  \label{tab:dataset_ablation}
  \centering
    \begin{tabular}{|p{3cm}|c||c|c|c|c||c|c|c|c|c|c|  }
     \hline
    \multirow{2}{*}{\textbf{Pre-training Dataset}} & \multirow{2}{*}{\textbf{Loss Type}} & 
    \multicolumn{4}{c||}{\textbf{Objective Metrics}} &
    \multicolumn{4}{c|}{\textbf{DNSMOS}} &
    \multicolumn{1}{c|}{\multirow{2}{*}{\textbf{PESQ}}} & 
    \multicolumn{1}{c|}{\multirow{2}{*}{\textbf{STOI}}}  \\ \cline{3-10} 

     \multicolumn{1}{|c|}{} & \multicolumn{1}{c||}{} & \multicolumn{1}{c|}{\textbf{SI-SDR}} & \multicolumn{1}{c|}{\textbf{SI-SDRi}} & \multicolumn{1}{c|}{\textbf{SDR}} & \multicolumn{1}{c||}{\textbf{SDRi}} & \multicolumn{1}{c|}{\textbf{OVRL}} & \multicolumn{1}{c|}{\textbf{SIG}} & \multicolumn{1}{c|}{\textbf{BAK}} & \multicolumn{1}{c|}{\textbf{p808}} & \multicolumn{1}{c|}{} & \multicolumn{1}{c|}{} \\ \hline
    
    \multirow{2}{*}{\textbf{AMUSE}} & Embedding & -1.16 & -1.16 & 1.10 & 0.94 & 1.80 & 2.12 & 2.99 & 2.58 & 1.80 & 0.81\\
     & Waveform & 2.8 & 2.7 & 4.9 & 4.7 & 1.81 & 2.04 & 3.38 & 2.55 & 2.06 & 0.79\\
    \hline
    
    \multirow{2}{*}{\textbf{Original}} & Embedding & -21.9 & -22.0 & -2.4 & -2.6 & 1.57 & 1.91 & 2.53 & 2.38 & 1.57 & 0.71\\
     & Waveform & 6.7 & 6.7 & 7.9 & 7.7 & 2.19 & 2.53 & 3.40 & 2.80 & 2.20 & 0.85\\
    \hline

    \multirow{2}{*}{\textbf{LibriTTS}} & Embedding & -3.4 & -3.3 & -0.1 & -0.2 & 1.89 & 2.26 & 2.98 & 2.62 & 1.75 & 0.80\\
     & Waveform & 1.0 & 1.1 & 3.0 & 2.9 & 1.82 & 2.03 & 3.43 & 2.60 & 1.76 & 0.75\\
    
    \hline
    \end{tabular}
\end{table*}
\subsection{Training Procedure}
All models were trained for 20 epochs on a V100 with 32GB RAM. The Adam optimizer was used with an initial learning rate of 1.5$e^{-4}$ and the LR scheduler was set to halve the learning rate with a patience of 2 after epoch 5. The models trained using the embedding loss were trained with a batch size of 20, while the those trained using the waveform loss were trained with a batch size of 3 due to memory limitations.

Following the settings of~\cite{yip2024towards}, the Codecformer separator consists of 16 transformer blocks with an input embedding size of 256. This is regardless of the embedding size of the neural audio codec used, which is either 1024 in the case of the original DAC and 512 for the ESPnet models, ensuring a fair comparison. The adapter layer in Codecformer was used to map the representations to the correct number of channels. 

\section{Results}
\subsection{Improvements to Training Speed and Computation}
\label{sec:fast-fast}
One of the key features of Codecformer-EL is the improvement in training speed and computational efficiently. We measure this in Table~\ref{tab:compandspeed} using Multiple and Accumulate operations (MACs) computed using the PyTorch-OpCounter\footnote{https://github.com/Lyken17/pytorch-OpCounter} as well as the number of hours required to train each epoch on the WSJ0-2mix dataset. We can see that Codecformer-EL trained 2.5x faster and with 1.9x fewer MACs compared to Codecformer. Additionally, it trained 6.8x faster than Sepformer and requires 97x fewer MACs. The MACs were calculated on 2 seconds of 8kHz audio data and a V100 GPU with 16GB of RAM utilizing the maximum possible batch size was used to calculate the training speed.

\renewcommand{\arraystretch}{1.2}
\begin{table}[h!]
  \caption{Computation and training speed of Codecformer-EL compared against Codecformer and Sepformer}
  \label{tab:compandspeed}
  \centering
    \begin{tabular}{ |c|c|c|  }
    \hline
    Model & GMACs & Training Time (h/epoch) \\
    \hline
    Sepformer~\cite{sepformer} & 77.3 & 2.7\\
    Codecformer~\cite{yip2024towards} & 1.5 & 1.0\\
    Codecformer-EL (Ours) & \textbf{0.8} & \textbf{0.4}\\
    \hline
    \end{tabular}
\end{table}

\vspace{-10pt}
\subsection{Comparison between Codecformer-EL and Codecformer}
Table~\ref{tab:modelComparisons} presents the separation performance of three neural audio codecs pre-trained on the AMUSE dataset, comparing our Codecformer-EL (embedding loss) method with the original Codecformer (waveform loss) approach.

A consistent pattern emerges: Codecformer models achieve higher scores on objective metrics (SI-SDR, SI-SDRi, SDR, SDRi), while Codecformer-EL models perform better on perceptual metrics, despite lower objective scores. This discrepancy highlights a known limitation of objective waveform-matching metrics for evaluating resynthesized speech \cite{shi2022discretizationresynthesisalternativemethod}.

The neural audio codec's decoder, trained with GAN-based loss, prioritizes generating perceptually natural speech over exact waveform reconstruction. This approach explains the disparity between objective and perceptual metric scores.
Across the models, we observe varying degrees of performance difference. DAC models show the smallest gap between methods, with embedding loss outperforming on some perceptual metrics. EnCodec models display the most dramatic difference, with embedding loss achieving the highest perceptual scores despite poor objective metrics. SoundStream models also demonstrate a clear advantage for embedding loss in perceptual metrics.

These results underscore the potential of Codecformer-EL for real-world applications where perceived quality is crucial. They also emphasize the importance of using perceptual quality metrics for evaluating resynthesized speech separation performance. The success of our method across different codec architectures demonstrates its versatility and potential for wide applicability in speech separation tasks.

\subsection{Comparison of speech separation performance across different codec pre-training datasets}
In Table~\ref{tab:dataset_ablation} we show the performance of the DAC model on speech separation after pre-training on different sets of data. In this analysis we see that regardless of the pre-training data, the DAC model performs similarly on the AMUSE and LibriTTS dataset, with embedding loss having a similar or better separation performance than waveform loss. However, in the model trained on the Original DAC dataset, the embedding method performs worse than the waveform loss and the Original DAC model with waveform loss performs the best out of all the models. This could potentially be due to the larger embedding size of 1024 compared to 512 in the other models. The embedding loss training method, due to its heavy reliance on the compressed representations, could be penalised more for the mismatch in dimension sizes between the transformer blocks and the DAC encoder. This could be further investigated in future work with ablation studies over different embedding sizes for the transformer in Codecformer-EL.

\subsection{Practical Considerations}
As discussed in Section~\ref{sec:fast-fast}, Codecformer-EL offers significant advantages in scenarios where neural audio codecs are already deployed or being considered. In these scenarios, because neural audio codecs are regenerating audio in a non-linear manner during decoding, it can cause distortions that result in a domain mismatch for signal-level speech separation tasks~\cite{yip2024towards}. 

One solution to this problem is to retrain or fine-tune a speech separation model on the outputs of neural audio codecs. However, this process can be slow and costly. On the other hand, the efficiency of Codecformer-EL makes it particularly valuable for rapid adaptation to new codecs, and is a more convenient approach compared to re-training separation models on new codec outputs at the signal level. 

By using compressed representations, Codecformer-EL reduces computational requirements for both training and inference, which could lead to faster deployment cycles and lower operational costs. The reduced computational demands also enhance scalability, potentially enabling deployment on a wider range of devices, including those with limited processing power.

\subsection{Limitations}
Despite its advantages, Codecformer-EL faces several limitations. Firstly, the upper bound of separation performance is inherently constrained by the quality and capabilities of the underlying neural audio codec. Since the model is trained on embeddings, the performance of Codecformer-EL heavily relies on the quality and generalizability of the pre-trained neural audio codec for both extracting mixture embeddings and providing the ground-truth labels. Therefore, suboptimal performance may result if the training data of the neural audio codec significantly differs from the data used to train Codecformer-EL.

Secondly, due to the generative nature of neural audio codecs, it is difficult to properly evaluate the performance of the separation output, limiting competitiveness with prior state-of-the-art results reported using signal-level loss. The variability introduced by the neural audio codec generation means that the output does not always align perfectly with original signal characteristics, despite having good perceptual quality. As observed in our results in Table~\ref{tab:modelComparisons} and~\ref{tab:dataset_ablation}, this creates a discrepancy between objective and perceptual metrics in the results, and poses challenges in directly comparing Codecformer-EL's performance with methods evaluated primarily on objective metrics.

Finally, working with learned representations in the compressed domain may reduce the interpretability of the separation process compared to methods operating directly on waveforms or spectrograms. Additionally, performance may vary across different neural audio codec architectures, potentially requiring codec-specific optimizations to achieve optimal results.
\vspace{-5pt}
\section{Conclusions}
In this paper, we introduced Codecformer-EL, a novel approach to speech separation that leverages neural audio codecs and employs an embedding-level loss function. Our method demonstrates significant improvements in computational efficiency and training speed compared to previous approaches, while maintaining comparable or superior perceptual quality in separated speech. We showed that Codecformer-EL is effective across multiple neural audio codecs and pre-training datasets, highlighting its versatility and potential for widespread application.

Our findings reveal an important dichotomy between objective and perceptual metrics in evaluating resynthesized speech. While objective metrics sometimes indicated poorer performance, perceptual quality metrics often showed comparable or superior results, underscoring the importance of using appropriate evaluation methods for this type of speech processing task.

Codecformer-EL's improved efficiency paves the way for more scalable speech separation models capable of handling real-world audio challenges, particularly in applications where audio compression is necessary for transmission, such as smartphones offloading computationally intensive tasks to cloud servers~\cite{yip2024towards}. However, it's crucial to acknowledge potential limitations, including the trade-off between compression and information preservation in codec embeddings. Highly compressed embeddings may lead to loss of fine-grained audio details, potentially affecting separation quality in complex acoustic environments~\cite{puvvada2024discrete}. Moreover, Codecformer-EL's performance is inherently tied to the quality and generalizability of the pre-trained neural audio codec, which may lead to suboptimal results when the codec's pre-training data significantly differs from the target separation task.

Future work could explore techniques to mitigate these limitations, such as fine-tuning the codec encoder for specific separation tasks or investigating adaptive compression levels based on input complexity. Additionally, research into the optimal balance between compression and separation quality could further enhance the practical applicability of this approach.

In conclusion, Codecformer-EL represents a significant step forward in efficient, high-quality speech separation, opening new avenues for both research and real-world applications in audio processing and communication technologies.

\section*{Acknowledgment}
This research is supported by the RIE2025 Industry Alignment Fund – Industry Collaboration Projects (IAF-ICP) (Award I2301E0026), administered by A*STAR, as well as supported by Alibaba Group and NTU Singapore. We would like to acknowledge Alibaba-NTU Joint Research Institute, Interdisciplinary Graduate Programme, Nanyang Technological University, Singapore

\newpage

\bibliographystyle{IEEEtran}
\bibliography{main}

\end{document}